\documentclass[aps,pra,twocolumn,showpacs,superscriptaddress]{revtex4}

\usepackage{graphicx}
\usepackage{latexsym}
\usepackage{amsmath}
\usepackage{amssymb}

\begin{document}
\title{Compact set of invariants characterizing graph states of up to eight qubits}
\author{Ad\'{a}n Cabello}
\email{adan@us.es}
\affiliation{Departamento de F\'{\i}sica Aplicada II,
Universidad de Sevilla, E-41012 Sevilla, Spain}
\author{Antonio J. L\'{o}pez-Tarrida}
\affiliation{Departamento de F\'{\i}sica Aplicada II, Universidad de
Sevilla, E-41012 Sevilla, Spain}
\author{Pilar Moreno}
\affiliation{Departamento de F\'{\i}sica Aplicada II, Universidad de
Sevilla, E-41012 Sevilla, Spain}
\author{Jos\'{e} R. Portillo}
\affiliation{Departamento de Matem\'{a}tica Aplicada I, Universidad
de Sevilla, E-41012 Sevilla, Spain}
\date{\today}




\begin{abstract}
The set of entanglement measures proposed by Hein, Eisert, and
Briegel for $n$-qubit graph states [Phys. Rev.~A {\bf 69}, 062311
(2004)] fails to distinguish between inequivalent classes under
local Clifford operations if $n \ge 7$. On the other hand, the set
of invariants proposed by van den Nest, Dehaene, and De Moor (VDD)
[Phys. Rev.~A {\bf 72}, 014307 (2005)] distinguishes between
inequivalent classes, but contains too many invariants (more than $2
\times 10^{36}$ for $n=7$) to be practical. Here we solve the
problem of deciding which entanglement class a graph state of $n \le
8$ qubits belongs to by calculating some of the state's intrinsic
properties. We show that four invariants related to those proposed
by VDD are enough for distinguishing between all inequivalent
classes with $n \le 8$ qubits.
\end{abstract}


\maketitle


\section{\label{Sec1}Introduction}


Graph states \cite{HEB04, HDERVB06} are fundamental in quantum
information, especially in quantum error correction
\cite{Gottesman96, SW02, Schlingemann02} and measurement-based
quantum computation \cite{RB01}. Graph states also play a
fundamental role in the study of entanglement. Two quantum states
have the same entanglement if they are equivalent under stochastic
local operations and classical communication (SLOCC). For $n=3$,
there are six classes under SLOCC \cite{DVC00}. For $n \ge 4$ the
number of classes under SLOCC is infinite and is specified by an
exponentially increasing number of parameters. However, if we focus
on graph states of $n < 27$ qubits, then the discussion becomes
simpler. On one hand, every two graph states which are SLOCC
equivalent are also equivalent under local unitary (LU) operations
\cite{VDM04b}. On the other hand, previous results suggest that, for
graph states of $n < 27$ qubits, the notion of LU equivalence and
local Clifford equivalence (LC equivalence) coincide. The
``LU$\Leftrightarrow$LC conjecture'' states that ``every two
LU-equivalent stabilizer states must also be LC equivalent.'' Ji
{\em et al.} proved that the LU$\Leftrightarrow$LC conjecture is
false \cite{JCWY07}. However, the LU$\Leftrightarrow$LC is true for
several classes of $n$ qubit graph states \cite{VDM05, ZCCC07} and
the simplest counterexamples to the conjecture are graph states of
$n=27$ qubits \cite{JCWY07}. Indeed, Ji {\em et al.} ``believe that
$27$ is the smallest possible size of counterexamples of
LU$\Leftrightarrow$LC.'' In this paper we assume that deciding
whether or not two graph states of $n < 27$ qubits have the same
entanglement is equivalent to deciding whether or not they are LC
equivalent.

The aim of this paper is to solve the following problem. Given an
$n$-qubit graph state with $n < 9$ qubits, decide which entanglement
class it belongs to just by examining some of the state's intrinsic
properties (i.e., without generating the whole LC class). The
solution to this problem is of practical importance. If one needs to
prepare a graph state $|G\rangle$ and knows that it belongs to one
specific class, then one can prepare $|G\rangle$ by preparing the
LC-equivalent state $|G'\rangle$ requiring the minimum number of
entangling gates and the minimum preparation depth of that class
(see \cite{HEB04, HDERVB06, CLMP08}) and then transform $|G'\rangle$
into $|G\rangle$ by means of simple one-qubit unitary operations.
The problem is that, so far, we do not know a simple set of
invariants which distinguishes between all classes of entanglement,
even for graph states with $n \le 7$ qubits.

The classification of graph states' entanglement has been achieved,
up to $n=7$ qubits, by Hein, Eisert, and Briegel (HEB) \cite{HEB04}
(see also \cite{HDERVB06}) and has recently been extended to $n=8$
qubits \cite{CLMP08}. The criteria for ordering the classes in
\cite{HEB04, HDERVB06, CLMP08} are based on several entanglement
measures: the minimum number of two-qubit gates required for the
preparation of a member of the class, the Schmidt measure for the
$n$-partite split (which measures the genuine $n$-party entanglement
of the class \cite{EB01}), and the Schmidt ranks for all bipartite
splits (or rank indexes \cite{HEB04, HDERVB06}). The problem is that
this set of entanglement measures fails to distinguish between
inequivalent classes (i.e., between different types of
entanglement). There is already an example of this problem in $n=7$:
none of these entanglement measures allows us to distinguish between
the classes 40, 42, and 43 in \cite{HEB04, HDERVB06}. A similar
problem occurs in $n=8$: none of these entanglement measures allows
us to distinguish between classes 110 and 111, between classes 113
and 114, and between classes 116 and 117 in \cite{CLMP08}.
Therefore, we cannot use these invariants for deciding which
entanglement class a given state belongs to. Reciprocally, if we
have such a set of invariants, then we can use it to unambiguously
label each of the classes.

Van den Nest, Dehaene, and De Moor (VDD) proposed a finite set of
invariants that characterizes all classes \cite{VDD05}. However,
already for $n=7$, this set has more than $2 \times 10^{36}$
invariants which are not explicitly calculated anywhere, so this set
is not useful for classifying a given graph state. Indeed, VDD
``believe that [their set of invariants] can be improved
---if not for all stabilizer states then at least for some
interesting subclasses of states'' \cite{VDD05}. Moreover, they
state that ``it is likely that only [some] invariants need to be
considered in order to recognize LC equivalence'' \cite{VDD05}, and
that ``it is not unlikely that there exist smaller complete lists of
invariants which exhibit less redundancies'' \cite{VDD05}. In this
paper we show that, if $n \le 8$, then four invariants are enough to
recognize the type of entanglement.

The paper is organized as follows. In Sec. \ref{Sec2} we introduce
some basic concepts of the graph state formalism and review some of
the results about the invariants proposed by VDD that will be useful
in our discussion. In Sec. \ref{Sec4} we present our results and in
Sec. \ref{Sec5} our conclusions.


\section{\label{Sec2}Basic concepts}


\subsection{Stabilizer}


The Pauli group $\mathcal{G}_{n}$ on $n$ qubits consists of all
$4\times4^{n}$ $n$-fold tensor products of the form $M=\alpha_{M}
M_{1}\bigotimes \cdots \bigotimes M_{n}$, where $\alpha_{M}\in\{\pm
1,\pm \imath\}$ is an overall phase factor and $M_{i}$ is either the
$2 \times 2$ identity matrix $\sigma_{0}=\openone$ or one of the
Pauli matrices $X=\sigma_{x}$, $Y=\sigma_{y}$, and $Z=\sigma_{z}$.

An $n$-qubit stabilizer $\mathcal{S}$ in the Pauli group is defined
as an Abelian subgroup of $\mathcal{G}_{n}$ which does not contain
the operator $-\openone$ \cite{NC00}. A stabilizer consists of
$2^{k}$ Hermitian (therefore, they must have real overall phase
factors $\pm 1$) $n$-qubit Pauli operators $s_{i}=\alpha_{i}
M_{1}^{(i)} \bigotimes \cdots \bigotimes M_{n}^{(i)}
\in\mathcal{G}_{n}, \, i=1,\ldots,2^{k}$ for some $k \leq n$. We
will call the operators $s_{i}$ stabilizing operators.

In group theory, a set of elements $\{g_{1},\ldots, g_{l}\}$ in a
group $G$ is said to generate the group $G$ if every element of $G$
can be written as a product of elements from $\{g_{1},\ldots,
g_{l}\}$. The notation $G=\langle g_{1}, \ldots, g_{l}\rangle$ is
commonly used to describe this fact, and the set $\{g_{1}, \ldots,
g_{l}\}$ is called the generator of $G$. The generator of an
$n$-qubit stabilizer $\mathcal{S}$ is a subset (not necessarily
unique) $\gamma_{\mathcal{S}}=\{g_{1},\ldots, g_{k}\}$, consisting
of $k\leq n$ independent stabilizing operators, such that
$\mathcal{S}=\langle \gamma_{\mathcal{S}}\rangle$. In this context,
independent means that no product of the form $g_{1}^{a_{1}} \cdots
g_{k}^{a_{k}}$, where $a_{i}\in\{0,1\}$ yields the identity except
when all $a_{i}=0$. As a consequence, removing any operator $g_{i}$
from the generator makes the generated group smaller.

By definition, given a stabilizer $\mathcal{S}$, the stabilizing
operators $s_{i}$ commute, so that they can be diagonalized
simultaneously and, therefore, share a common set of eigenvectors
that constitute a basis of the so-called vector space
$V_{\mathcal{S}}$ stabilized by $\mathcal{S}$. The vector space
$V_{\mathcal{S}}$ is of dimension $2^{q}$ when
$|\gamma_{\mathcal{S}}|=n-q$. Remarkably, if $|\mathcal{S}|=2^{n}$,
then there exists a unique common eigenstate $|\psi\rangle$ on $n$
qubits with eigenvalue $1$, such that
$s_{i}|\psi\rangle=|\psi\rangle$ for every stabilizing operator
$s_{i}\in\mathcal{S}$. Such a state $|\psi\rangle$ is called a
stabilizer state because it is the only state that is fixed
(stabilized) by every operator of the stabilizer $\mathcal{S}$.

Graph states are a special kind of stabilizer states (with $k=n$)
associated with graphs. It has been demonstrated that every
stabilizer state is equivalent under local complementation (defined
below) to some (generally non unique) graph state
\cite{Schlingemann02}.


\subsection{Graph state}


A $n$-qubit graph state $|G\rangle$ is a pure state associated to a
graph $G (V, E)$ consisting of a set of $n$ vertices
$V=\{1,\ldots,n\}$ and a set of edges $E$ connecting pairs of
vertices, $E \subset V \times V$. Each vertex represents a qubit.
The graph $G$ provides a mathematical characterization of
$|G\rangle$. The graph state $|G\rangle$ associated to the graph $G$
is the unique $n$-qubit state fulfilling
\begin{equation}
g_i |G\rangle= |G\rangle, \quad{\rm for}\quad i=1,\ldots,n,
\label{graphdef}
\end{equation}
where $g_i$ are the generators of the state's stabilizer group,
defined as the set $\{s_j\}_{j=1}^{2^n}$ of all products of the
generators. $g_i$ is the generator operator associated to the vertex
$i$, defined by
\begin{equation}
g_i:= X^{(i)}\bigotimes\nolimits_{(i,j)\in E} Z^{(j)},
\label{stabdef}
\end{equation}
where the product is extended to those vertices $j$ which are
connected with $i$ and $X^{(i)}$ ($Z^{(i)}$) denotes the Pauli
matrix $\sigma_x$ ($\sigma_z$) acting on the $i$th qubit.


\subsection{\label{localcomplementation}Local complementation}


Two $n$-qubit states, $|\phi\rangle$ and $|\psi\rangle$, have the
same $n$-partite entanglement if and only if there are $n$ one-qubit
unitary transformations $U_i$, such that $|\phi\rangle =
\bigotimes_{i=1}^n U_i |\psi\rangle$. If these one-qubit unitary
transformations belong to the Clifford group, then the two states
are said to be LC equivalent. VDD found that the successive
application of a transformation with a simple graphical description
is enough to generate the complete equivalence class of graph states
under local unitary operations within the Clifford group (hereafter
simply referred to as class or orbit) \cite{VDM04}. This simple
transformation is local complementation.

On the stabilizer, local complementation on the qubit $i$ induces
the map $Y^{(i)} \mapsto Z^{(i)}, Z^{(i)} \mapsto -Y^{(i)}$ on the
qubit $i$ and the map $X^{(j)} \mapsto -Y^{(j)}, Y^{(j)} \mapsto
X^{(j)}$ on the qubits $j$ connected with $i$ \cite{HDERVB06}. On
the generators, local complementation on the qubit $i$ maps the
generators $g^{\rm{old}}_j$, with $j$ connected with $i$ to
$g^{\rm{new}}_j g^{\rm{new}}_{i}$.

\begin{figure}[htb]
\centerline{\includegraphics[width=0.75
\columnwidth]{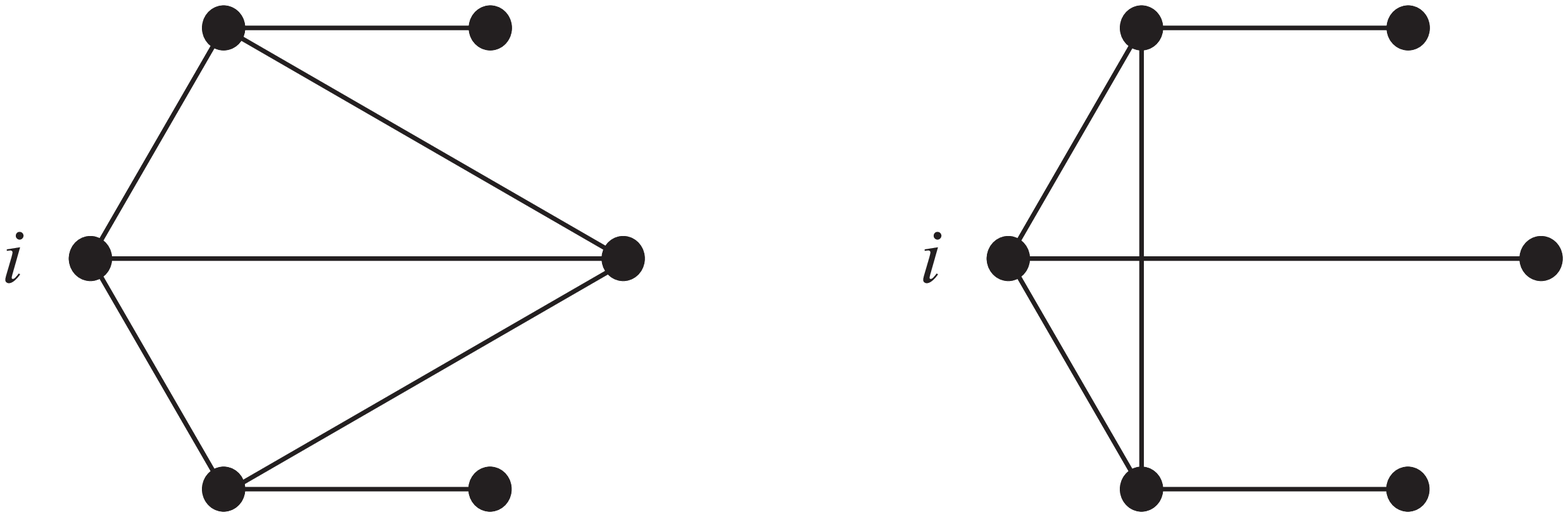}} \caption{\label{Fig1}
Graphical effect of local complementation on qubit $i$. Local
complementation on qubit $i$ on the graph on the left (right) leads
to the graph on the right (left).}
\end{figure}

Graphically, local complementation on qubit $i$ acts as follows.
Those vertices connected with $i$ which were connected from each
other become disconnected from each other and vice versa. An example
is in Fig.~\ref{Fig1}.


Using local complementation, one can generate the orbits of all
LC-inequivalent $n$-qubit graph states. There are $45$ orbits for $n
\le 7$ \cite{HEB04, HDERVB06} and $101$ orbits for $n = 8$
\cite{CLMP08}.


\subsection{Supports and LC invariants related to supports}


Let $|\psi\rangle$ be a stabilizer state and
$\mathcal{S}(|\psi\rangle)$ the corresponding stabilizer. Given a
stabilizing operator $s_{i}=\alpha_{i} M_{1}^{(i)} \bigotimes \cdots
\bigotimes M_{n}^{(i)}$, its support $\mbox{supp}(s_{i})$ is the set
of all $j\in\{1,\ldots,n\}$ such that $M_{j}^{(i)}$ differs from the
identity. Therefore, the support of $s_{i}$ is the set of the labels
of the qubits on which the action of the Pauli matrices is non
trivial (i.e., there is a $X$, $Y$, or $Z$ Pauli matrix acting on
the qubit). Notice that the support is preserved under the maps
induced on the stabilizer by local complementation (see
Sec.~\ref{localcomplementation}).

Let $\omega \subseteq \{1,\ldots,n\}$ be the support of a
stabilizing operator $s_{i}$, $\mbox{supp}(s_{i})=\omega$. The
weight of the operator $s_{i}$ is the cardinality of its support,
$|\omega|$. The identity operator $\openone \bigotimes \cdots
\bigotimes \openone$, which is always present in a stabilizer due to
the underlying group structure, fulfills $\omega=\{\emptyset\}$ and,
therefore, is of weight zero.

The set of operators $\{s_{i}\}_{i=1}^{2^{k}}$ of a stabilizer
$\mathcal{S}$ can be classified into equivalence classes according
to their supports, defining a partition in the stabilizer. We will
say that two stabilizing operators $s_{i}$ and $s_{j}$ of
$\mathcal{S}$ belong to the same equivalence class $[\omega]$ if
they have the same support $\omega$, i.e., $\mbox{supp}(s_{i})=
\mbox{supp}(s_{j})=\omega$. We denote by $A_{\omega}(|\psi\rangle)$
the number of elements (stabilizing operators) $s_{i} \in
\mathcal{S}(|\psi\rangle)$ with $\mbox{supp}(s_{i})= \omega$. In
other words, $A_{\omega}(|\psi\rangle)$ is the cardinality of the
equivalence class $[\omega]$. Since any graph state $|G\rangle$ is a
special type of stabilizer state, these definitions can also be
applied to them.


\subsection{\label{Sec3}Invariants of Van den Nest,
Dehaene, and De Moor}


The following theorem is a key result obtained by VDD in
\cite{VDD05} that presents a finite set of invariants which
characterizes the LC equivalence class of any stabilizer state
(i.e., functions that remain invariant under the action of all local
Clifford transformations). We have chosen an adapted formulation of
the theorem to group multiplication involving Pauli operators [see
Eq.~\eqref{supp cond}], slightly different from VDD's original
notation, which is based on the well-known equivalent formulation of
the stabilizer formalism in terms of algebra over the field
$\mathbb{F}_{2}={\rm GF}(2)$, where arithmetic is performed modulo
$2$ and each stabilizing operator is identified with a
$2n$-dimensional binary index operator.

{\em Theorem~1.} Let $|\psi\rangle$ be a stabilizer state on $n$
qubits corresponding to a stabilizer $\mathcal{S}_{|\psi\rangle}$.
Let $r \in \mathbb{N}_0$ and consider subsets $\omega_{k},
\omega_{kl}\subseteq\{1,\ldots,n\}$ for every $k, l \in
\{1,\ldots,r\}$, with $k<l$. Denote $\Omega:=(\omega_{1},
\omega_{2},\ldots,\omega_{12},\omega_{13},\ldots)$ and let
$\mathcal{T}_{n,r}^{\Omega}(|\psi\rangle)$ be the set consisting of
all tuples
$(s_{1},\ldots,s_{r})\in\mathcal{S}_{|\psi\rangle}\times\ldots\times\mathcal{S}_{|\psi\rangle}$
satisfying
\begin{equation}\label{supp cond}
\mbox{supp}(s_{k})=\omega_{k},\;\mbox{supp}(s_{k}
s_{l})=\omega_{kl}.
\end{equation}
Then, (i) $|\mathcal{T}_{n,r}^{\Omega}(|\psi\rangle)|$ is LC
invariant and (ii) the LC equivalence class of $|\psi\rangle$ is
completely determined by the values of all invariants
$|\mathcal{T}_{n,n}^{\Omega}(|\psi\rangle)|$ (i.e., where $r=n$).

VDD provide another family of support-related invariants, based on a
second theorem with the same formulation than the one above, except
for the substitution of conditions \eqref{supp cond} by new
constraints
\begin{equation}\label{supp cond2}
\mbox{supp}(s_{k})\subseteq\omega_{k},\;\mbox{supp}(s_{k}
s_{l})\subseteq\omega_{kl}.
\end{equation}
These new LC invariants are the dimensions of certain vector spaces
and, in principle, are more manageable from a computational point of
view because they involve the generator matrix of the stabilizer and
rank calculation. Nevertheless, we will focus our attention on the
first family of invariants, since they suffice to solve the problem
we address in this paper with no extra computational effort. To
resort to the second family would be justified in case we had to use
invariants with a high $r$ value to achieve LC discrimination among
graph states up to eight qubits. We refer the reader to
Ref.~\cite{VDD05} for a proof of Theorem~1 and the extension to the
second family of LC invariants.

The invariants of Theorem~1 are the cardinalities of certain subsets
$\mathcal{T}_{n,r}^{\Omega}(|\psi\rangle)$ of
$\mathcal{S}_{|\psi\rangle}\cdots\mathcal{S}_{|\psi\rangle}$, which
are defined in terms of simple constraints \eqref{supp cond} on the
supports of the stabilizing operators. VDD pointed out that, for
$r=1$, these invariants count the number of operators in the
stabilizer with a prescribed support. Therefore, fixing $r=1$, for
every possible support $\omega_{k}\subseteq\{1,\ldots,n\}$, there is
an invariant
\begin{equation}\label{inv r1}
|\{s\in\mathcal{S}_{|\psi\rangle}|\mbox{supp}(s)=\omega_{k}\}|.
\end{equation}
That is, the invariants for $r=1$ are the
$A_{\omega_{k}}(|\psi\rangle)$, i.e., the cardinalities of the
equivalence classes $[\omega_{k}]$ of the stabilizer. The number of
possible supports in a stabilizer of an $n$-qubit state is equal to
$2^{n}$ and, therefore, there are $2^{n}$ VDD's invariants for
$r=1$. Many of them could be equal to zero. In fact, when dealing
specifically with graph states, it can be easily seen that
$A_{\omega_{k}}(|\psi\rangle)=0$ when referred to supports
fulfilling $|\omega_{k}|=1$ because stabilizing operators of
weight~1 are not present in the stabilizer of a graph state due to
the inherent connectivity of the graphs associated to the states
that rules out isolated vertices.

On the other hand, VDD consider the invariants
$A_{\omega_{k}}(|\psi\rangle)$ as ``local versions'' of the
so-called weight distribution of a stabilizer, a concept frequently
used in classical and quantum coding theory. For $r\ge 2$, the new
series of invariants involve $r$-tuples of stabilizing operators and
their corresponding supports and constitute a generalization of the
weight distribution. Let us denote
\begin{equation}\label{suma pesos}
A_{d}(|\psi\rangle)=\sum_{\omega,
|\omega|=d}A_{\omega}(|\psi\rangle),
\end{equation}
the number of stabilizing operators with weight equal to $d$.
According to this notation, the weight distribution of a stabilizer
is the $(n+1)$-tuple
\begin{equation}\label{distribW}
W_{|\psi\rangle}=\{A_{d}(|\psi\rangle)\}_{d=0}^{n}.
\end{equation}

In principle, $W_{|\psi\rangle}$ could be a compact way to present
the whole information about the invariants
$A_{\omega}(|\psi\rangle)$, i.e., VDD's invariants with $r=1$. This
question will be addressed later.

In order to clarify the content of VDD's theorem, let us briefly
discuss the way it works when applied to a particular graph state.
We have chosen the three-qubit linear cluster state,
$|\rm{LC}_{3}\rangle$, because of its simplicity, combined with a
sufficient richness in the stabilizer structure. Table~\ref{TableI}
shows the stabilizer of $|\rm{LC}_{3}\rangle$ with its eight
stabilizing operators, $\{s_1,\ldots,s_8\}$. Three of them
($s_1=g_1$, $s_2=g_2$, and $s_3=g_3$) constitute a generator.
$|\rm{LC}_{3}\rangle$ is a three-qubit graph state, so there are
$2^{3}=8$ possible supports ($8$ being the number of subsets in the
set $\{1,2,3\}$):
\begin{equation}\label{listsupp}
\{\emptyset\}, \{1\},\{2\},\{3\},\{1,2\},\{1,3\},\{2,3\},\{1,2,3\}.
\end{equation}


\begin{table}[htb]
\caption{\label{TableI}Stabilizer and supports for the
$|\rm{LC}_{3}\rangle$.}
\begin{ruledtabular}
{\begin{tabular}{cccc} Stabilizing operators & & Support & Weight \\
\hline \hline
$XZ\openone$ & $s_1=g_1$ & $\{1,2\}$ & $2$ \\
$ZXZ$ & $s_2=g_2$ & $\{1,2,3\}$ & $3$ \\
$\openone ZX$ & $s_3=g_3$ & $\{2,3\}$ & $2$ \\
$\openone \openone \openone$ & $s_4=g_1 g_1$ & $\{\emptyset\}$ & $0$ \\
$YYZ$ & $s_5=g_1 g_2$ & $\{1,2,3\}$ & $3$ \\
$X\openone X$ & $s_6=g_1 g_3$ & $\{1,3\}$ & $2$ \\
$ZYY$ & $s_7=g_2 g_3$ & $\{1,2,3\}$ & $3$ \\
$-YXY$ & $s_8=g_1 g_2 g_3$ & $\{1,2,3\}$ & $3$ \\
\end{tabular}}
\end{ruledtabular}
\end{table}


VDD's invariants for $r=1$. In this case, $\Omega=(\omega_{1})$. By
$\Omega$, we denote each of all the possible ways to choose a single
support $\omega_{1}$, so there are eight choices for $\Omega$, which
are those listed in Eq.~\eqref{listsupp}. Given a particular choice
of $\Omega=(\omega_{1})$, the set
$\mathcal{T}_{n,1}^{\Omega}(|\rm{LC}_{3}\rangle)$ contains all
stabilizing operators $s_1$ for the $|\rm{LC}_{3}\rangle$ fulfilling
\begin{equation}\label{supp cond1}
\mbox{supp}(s_1)=\omega_1,
\end{equation}
so, as a matter of fact,
$\mathcal{T}_{n,1}^{\Omega}(|\rm{LC}_{3}\rangle)$ is the equivalence
class $[\omega_1]$ associated to the support $\omega_1$. Only five
out of the eight possible supports are in fact present in the
stabilizer of the $|\rm{LC}_{3}\rangle$ (see the column ``Support''
in Table~\ref{TableI}) and, therefore, we can distinguish between
five non empty equivalence classes $[\omega]$. According to VDD's
theorem, the LC invariants for $r=1$,
$|\mathcal{T}_{n,1}^{\Omega}(|\rm{LC}_{3}\rangle)|$, are the
cardinalities $A_{\omega}(|\rm{LC}_{3}\rangle)$ of such equivalence
classes $[\omega]$, namely,
\begin{subequations}
\begin{align}
A_{\{\emptyset\}}(|\rm{LC}_{3}\rangle)&=1, \label{10a} \\
A_{\{1\}}(|\rm{LC}_{3}\rangle)&=0,\\
A_{\{2\}}(|\rm{LC}_{3}\rangle)&=0,\\
A_{\{3\}}(|\rm{LC}_{3}\rangle)&=0,\\
A_{\{1,2\}}(|\rm{LC}_{3}\rangle)&=1,\\
A_{\{1,3\}}(|\rm{LC}_{3}\rangle)&=1,\\
A_{\{2,3\}}(|\rm{LC}_{3}\rangle)&=1,\\
A_{\{1,2,3\}}(|\rm{LC}_{3}\rangle)&=4. \label{10h}
\end{align}
\end{subequations}

VDD's invariants for $r=2$. In this case, $\Omega=(\omega_{1},
\omega_{2}; \omega_{12})$. By $\Omega$ we denote each of all the
possible different ways to choose two supports $ \omega_{1}$,
$\omega_{2}$, and then a third support $\omega_{12}$. Let $M=2^n$ be
the number of possible supports and $n$ being the number of qubits.
On one hand, there are $\binom{M}{2}$ different combinations of two
supports $(\omega_{1}, \omega_{2})$, plus $M$ couples of the form
$(\omega_{1}, \omega_{2}=\omega_{1})$. On the other hand, there are
$M$ possible choices for $\omega_{12}$. As a consequence, there are
$M\left[M+\binom{M}{2}\right]$ ways to choose $\Omega$. For $n=3$,
this number is $288$. Given a particular choice of
$\Omega=(\omega_{1}, \omega_{2}; \omega_{12})$, the set
$\mathcal{T}_{n,2}^{\Omega}(|\rm{LC}_{3}\rangle)$ contains all the
two-tuples of the stabilizing operators $(s_1,s_2)$ of the
$|\rm{LC}_{3}\rangle$ fulfilling
\begin{equation}\label{supp cond ej}
\mbox{supp}(s_1)=\omega_1,\;\mbox{supp}(s_2)=\omega_2,\;
\mbox{supp}(s_1 s_2)=\omega_{12}.
\end{equation}
Many of these sets $\mathcal{T}_{n,2}^{\Omega}(|\rm{LC}_{3}\rangle)$
could be empty because the stabilizer fails to fulfill any of the
conditions \eqref{supp cond ej}. The cardinalities of the $288$
sets, $|\mathcal{T}_{n,2}^{\Omega}(|\rm{LC}_{3}\rangle)|$, are the
VDD's invariants that we are interested in. For instance, if we
choose $\Omega=(\omega_{1}, \omega_{2}; \omega_{12})$ such that
\begin{equation}\label{example}
\omega_1=\{1,2\},\;\omega_2=\{1,2,3\},\;\omega_{12}=\{1,2,3\},
\end{equation}
then, according to the information in Table~\ref{TableI}, there is
only one operator with support $\omega_1$, namely, $s_1$, and four
operators with support $\omega_2$: $s_2$, $s_5$, $s_7$, and $s_8$.
We obtain $\mathcal{T}_{n,2}^{\Omega}(|\rm{LC}_{3}\rangle)=\{(s_1,
s_2), (s_1, s_5), (s_1, s_7), (s_1, s_8)\}$ because these four
two-tuples verify conditions \eqref{supp cond ej}, and the value of
the corresponding VDD's invariant is the cardinality of the set,
$|\mathcal{T}_{n,2}^{\Omega}(|\rm{LC}_{3}\rangle)|=4$.

Another example. If we choose $\Omega$ such that $\omega_1=\{1,2\}$,
$\omega_2=\{2,3\}$, and $\omega_{12}=\{1,2,3\}$, then the VDD's
invariant is $|\mathcal{T}_{n,2}^{\Omega}(|\rm{LC}_{3}\rangle)|=0$,
because the two-tuple $(s_1, s_3)$ defined by the supports
$\omega_1$ and $\omega_2$ fulfills $s_1 s_3 =s_6$ and $s_6$ does not
match with support $\omega_{12}$.


\section{\label{Sec4}Results}


The number of VDD's invariants for an $n$-qubit graph state grows
very rapidly with $r$ (and, of course, with $n$). If $n=3$, there
are eight invariants for $r=1$ and $288$ invariants for $r=2$. For
an eight-qubit graph state, there are $256$ invariants for $r=1$ and
$8421376$ for $r=2$. Obviously, the problem of calculating all the
VDD's invariants for graph states up to eight qubits becomes
completely unfeasible if there are no restrictions on $r$. The total
number of VDD's invariants for a given $n$-qubit graph state, and
all possible values of $r$, is $M+\sum_{r=2}^n C'(M,r) C'(M,P)$,
where $M=2^n$, $P=\binom{r}{2}$, and $C'(M,r)$ denotes the
combinations with repetition of $M$ elements choose $r$. For $n=7$,
this formula gives $2.18 \times 10^{36}$; for $n=8$, it gives $1.88
\times 10^{53}$.

How many of them are needed to distinguish between all LC
equivalence classes? VDD stated that ``the LC equivalence class of
$|\psi\rangle$ is completely determined by the values of all
invariants $|\mathcal{T}_{n,n}^{\Omega}(|\psi\rangle)|$ (i.e., where
$r=n$)'' \cite{VDD05}. However, this number [i.e., $C'(M,r) C'(M,P)$
with $r=n$] is still too large to be practical. For $n=7$ is $2.18
\times 10^{36}$ and for $n=8$ is $1.88 \times 10^{53}$ (i.e., most
of the invariants correspond to the case $r=n$). We are interested
in the minimum value of $r$ that yields a series of invariants
sufficient to distinguish between all the $146$ LC equivalence
classes of graph states up to $n=8$ qubits. In Ref.~\cite{VDD05},
the authors point out that there are examples of equivalence classes
in stabilizer states which are characterized by invariants of small
$r$; for instance, those equivalent to GHZ states. In addition, they
remark that a characterization based on small $r$ values could be
feasible, at least for some interesting subclasses or subsets of
stabilizer states. We have calculated the VDD's invariants for $r=1$
for the $146$ LC equivalence classes of graph states with up to
$n=8$ qubits. This implies calculating the cardinalities
$A_{\omega}(|\psi\rangle)$ of the corresponding equivalence classes
$[\omega]$ of the $146$ representatives of the LC equivalence
classes, $30060$ invariants in total (since there are 1, 1, 2, 4,
11, 26, and 101 classes of two-, three-, four-, five-, six-, seven-,
and eight-qubit graph states, respectively, and the number of VDD's
invariants with $r=1$ is $2^n$ for each class). Our results confirm
the conjecture that invariants with $r=1$ are enough for
distinguishing between the $146$ LC equivalence classes for graph
states up to eight qubits. It is therefore unnecessary to resort to
families of VDD's invariants
$|\mathcal{T}_{n,r}^{\Omega}(|\psi\rangle)|$ with $r \geq 2$.

Our goal is not to show the values of these $30060$ invariants but
to compress all this information and construct simple invariants
from it. However, in order to do it properly, some requirements
should be fulfilled. (I) The compacted information must be
unambiguous and easily readable. (II) The compacted information must
be LC invariant. (III) The compacted information concerning
different LC equivalence classes must still distinguish between any
of them.

Following the comments of VDD in Ref.~\cite{VDD05} about considering
the invariants $A_{\omega}(|\psi\rangle)$ as ``local versions'' of
the weight distribution $W_{|\psi\rangle}$ of a stabilizer, we have
calculated $W_{|\psi\rangle}$ for the $146$ LC classes of
equivalence, according to definition \eqref{distribW}. It can easily
be seen that, if $A_{\omega}(|\psi\rangle)$ is LC invariant, then
$W_{|\psi\rangle}$ is also LC invariant and permits a compact way to
compress the information of the invariants
$A_{\omega}(|\psi\rangle)$. Unfortunately, $W_{|\psi\rangle}$ is not
able to distinguish between any two LC classes of equivalence.
Table~\ref{TableII} shows that the weight distribution fails to
distinguish between LC classes starting from $n=6$. Graph states
with labels $13$ and $15$ in Refs.~\cite{HEB04, HDERVB06} have the
same weight distribution and this degeneration increases as the
number of qubits grows, as we have checked out calculating
$W_{|\psi\rangle}$ for all graph states up to eight qubits.

\begin{table}[htb]
\caption{\label{TableII}Weight distribution for graph states up to
six qubits.}
\begin{ruledtabular}
{\begin{tabular}{cccccccc} Graph state & $A_0$ & $A_1$ & $A_2$ & $A_3$ & $A_4$ & $A_5$ & $A_6$ \\
\hline \hline
$1$ & $1$ & $0$ & $3$ & & & & \\
$2$ & $1$ & $0$ & $3$ & $4$ & & & \\
$3$ & $1$ & $0$ & $6$ & $0$ & $9$ & & \\
$4$ & $1$ & $0$ & $2$ & $8$ & $5$ & & \\
$5$ & $1$ & $0$ & $10$ & $0$ & $5$ & $16$ & \\
$6$ & $1$ & $0$ & $4$ & $6$ & $11$ & $10$ & \\
$7$ & $1$ & $0$ & $2$ & $8$ & $13$ & $8$ & \\
$8$ & $1$ & $0$ & $0$ & $10$ & $15$ & $6$ & \\
$9$ & $1$ & $0$ & $15$ & $0$ & $15$ & $0$ & $33$ \\
$10$ & $1$ & $0$ & $7$ & $8$ & $7$ & $24$ & $17$ \\
$11$ & $1$ & $0$ & $6$ & $0$ & $33$ & $0$ & $24$ \\
$12$ & $1$ & $0$ & $4$ & $8$ & $13$ & $24$ & $14$ \\
$13$ & $1$ & $0$ & $3$ & $8$ & $15$ & $24$ & $13$ \\
$14$ & $1$ & $0$ & $2$ & $8$ & $17$ & $24$ & $12$ \\
$15$ & $1$ & $0$ & $3$ & $8$ & $15$ & $24$ & $13$ \\
$16$ & $1$ & $0$ & $3$ & $0$ & $39$ & $0$ & $21$ \\
$17$ & $1$ & $0$ & $1$ & $8$ & $19$ & $24$ & $11$ \\
$18$ & $1$ & $0$ & $0$ & $8$ & $21$ & $24$ & $10$ \\
$19$ & $1$ & $0$ & $0$ & $0$ & $45$ & $0$ & $18$ \\
\end{tabular}}
\end{ruledtabular}
\end{table}


Therefore, we must look for a way to compress the information about
the invariants $A_{\omega}(|\psi\rangle)$, which satisfies
(I)--(III). The fact that the stabilizing operators of a stabilizer
can be classified into equivalence classes according to their
supports (equivalence classes $[\omega]$), and that the
cardinalities of such classes $[\omega]$ are the invariants
$A_{\omega}(|\psi\rangle)$, leads us to introduce two definitions.
Two classes $[\omega_1]$ and $[\omega_2]$ are equipotent if and only
if both have the same cardinality, i.e.,
$A_{\omega_1}(|\psi\rangle)=A_{\omega_2}(|\psi\rangle)$, regardless
of whether their stabilizing operators have different weights
$|\omega_1|\neq|\omega_2|$ or not. It is clear that the number of
equipotent equivalence classes $[\omega]$ for a given cardinality
$A_{\omega}(|\psi\rangle)$ is LC invariant. We will call it the
$A_{\omega}$ multiplicity (or $A_{\omega}$ potency) and denote it by
$M(A_\omega)$. For instance, if we take a look at the list of
invariants $A_{\omega}(|\rm{LC}_{3}\rangle)$ [see
Eqs.~\eqref{10a}--\eqref{10h}] we find that the value $0$ appears
three times (so there are three equivalence classes $[\omega]$ with
that cardinality), and then $M(0)=3$. Using this criterion, $M(1)=4$
and $M(4)=1$ for the $|\rm{LC}_{3}\rangle$.

If we tabulate the values of $A_{\omega}(|\psi\rangle)$ together
with the corresponding values of $M(A_\omega)$, we obtain a
two-index compact information, which is LC invariant and, more
importantly, LC discriminant, as required. The results are shown in
Tables~\ref{TableIII}--\ref{TableV}.

In Table~\ref{TableV} we can see that four numbers are enough to
distinguish between all classes of graph states with $n=8$ qubits:
the multiplicities of the values $0$, $1$, $3$, and $4$. Indeed, in
Tables~\ref{TableIII} and \ref{TableIV} we see that these four
numbers are enough to distinguish between all classes of graph
states with $n \le 8$ qubits.


\section{\label{Sec5}Conclusions}


We have shown that, to decide which entanglement class a graph state
of $n \le 8$ qubits belongs to, it is enough to calculate four
quantities. These four LC invariants characterize any LC class of $n
\le 8$ qubits.

This result solves a problem raised in the classification of graph
states of $n \le 8$ qubits developed in Refs.~\cite{HEB04, HDERVB06,
CLMP08}. A compact set of invariants that characterize all
inequivalent classes of graph states with a higher number of qubits
can be obtained by applying the same strategy. This can be done
numerically up to $n=12$, a number of qubits beyond the present
experimental capability in the preparation of graph states
\cite{Gao08}.

We have also shown that the conjecture \cite{Bouchet93} that the
list of LC invariants given in Eq.~\eqref{supp cond2} is sufficient
to characterize the LC equivalence classes of all stabilizer states,
which is not true in general \cite{HEB04}, is indeed true for graph
states of $n \le 8$ qubits. Moreover, we have shown that, for graph
states of $n \le 8$ qubits, the list of LC invariants given in
Eq.~\eqref{inv r1}, which is more restrictive than the list given in
Eq.~\eqref{supp cond2}, is enough. This solves a problem suggested
in \cite{VDD05}, regarding the possibility of characterizing special
subclasses of stabilizer states using subfamilies of invariants.


\section*{Acknowledgements}


The authors thank H. J. Briegel, O. G{\"u}hne, M. Hein, and M. Van
den Nest, for their help. A.C., A.J.L., and P.M. acknowledge support
from Projects No.~P06-FQM-02243, No.~FIS2008-05596, and
No.~PAI-FQM-0239. J.R.P. acknowledges support from Projects
No.~P06-FQM-01649, No.~MTM2008-05866-C03-01, and No.~PAI-FQM-0164.


\begin{table}[t]
\caption{\label{TableIII}Invariants for the $n$-qubit graph states
with $3 \le n \le 6$. Notation: value$_{\rm multiplicity}$. The
numeration of the classes is the one in \cite{HEB04, HDERVB06}.}
\begin{ruledtabular}
{\begin{tabular}{cc} No. & Invariants \\
\hline
1 & $0_{2}$, $1_{1}$, $3_{1}$ \\
2 & $0_{3}$, $1_{4}$, $4_{1}$ \\
3 & $0_{8}$, $1_{7}$, $9_{1}$ \\
4 & $0_{8}$, $1_{3}$, $2_{4}$, $5_{1}$ \\
5 & $0_{15}$, $1_{16}$, $16_{1}$ \\
6 & $0_{18}$, $1_{8}$, $2_{3}$, $4_{2}$, $10_{1}$ \\
7 & $0_{17}$, $1_{7}$, $2_{6}$, $5_{1}$, $8_{1}$ \\
8 & $0_{15}$, $1_{11}$, $3_{5}$, $6_{1}$ \\
9 & $0_{32}$, $1_{31}$, $33_{1}$ \\
10 & $0_{38}$, $1_{15}$, $2_{8}$, $8_{2}$, $17_{1}$ \\
11 & $0_{41}$, $1_{16}$, $4_{6}$, $24_{1}$ \\
12 & $0_{38}$, $1_{14}$, $2_{7}$, $4_{1}$, $5_{2}$, $8_{1}$, $14_{1}$ \\
13 & $0_{42}$, $1_{6}$, $2_{8}$, $4_{6}$, $5_{1}$, $13_{1}$ \\
14 & $0_{37}$, $1_{12}$, $2_{8}$, $3_{4}$, $6_{2}$, $12_{1}$ \\
15 & $0_{42}$, $1_{12}$, $4_{6}$, $5_{3}$, $13_{1}$ \\
16 & $0_{44}$, $1_{4}$, $2_{12}$, $5_{3}$, $21_{1}$ \\
17 & $0_{34}$, $1_{18}$, $2_{6}$, $3_{1}$, $5_{4}$, $11_{1}$ \\
18 & $0_{33}$, $1_{21}$, $3_{3}$, $4_{6}$, $10_{1}$ \\
19 & $0_{47}$, $1_{1}$, $3_{15}$, $18_{1}$ \\
\end{tabular}}
\end{ruledtabular}
\end{table}


\begin{table}[t]
\caption{\label{TableIV}Invariants for the seven-qubit graph states.
Notation: value$_{\rm multiplicity}$. The numeration of the classes
is the one in \cite{HEB04, HDERVB06}.}
\begin{ruledtabular}
{\begin{tabular}{cc} No. & Invariants \\
\hline
20 & $0_{63}$, $1_{64}$, $64_{1}$ \\
21 & $0_{78}$, $1_{32}$, $2_{15}$, $16_{2}$, $34_{1}$ \\
22 & $0_{84}$, $1_{32}$, $4_{8}$, $8_{3}$, $40_{1}$ \\
23 & $0_{77}$, $1_{31}$, $2_{15}$, $8_{3}$, $17_{1}$, $26_{1}$ \\
24 & $0_{87}$, $1_{25}$, $4_{8}$, $5_{6}$, $16_{1}$, $25_{1}$ \\
25 & $0_{92}$, $1_{12}$, $2_{8}$, $4_{7}$, $5_{4}$, $8_{4}$, $20_{1}$ \\
26 & $0_{87}$, $1_{16}$, $2_{14}$, $4_{7}$, $8_{1}$, $10_{2}$, $28_{1}$ \\
27 & $0_{80}$, $1_{25}$, $2_{11}$, $3_{3}$, $4_{3}$, $5_{3}$, $8_{1}$, $14_{1}$, $23_{1}$ \\
28 & $0_{85}$, $1_{15}$, $2_{16}$, $4_{3}$, $6_{7}$, $9_{1}$, $18_{1}$ \\
29 & $0_{87}$, $1_{12}$, $2_{15}$, $4_{9}$, $5_{3}$, $13_{1}$, $22_{1}$ \\
30 & $0_{80}$, $1_{21}$, $2_{12}$, $3_{6}$, $4_{1}$, $5_{4}$, $8_{3}$, $17_{1}$ \\
31 & $0_{86}$, $1_{28}$, $4_{3}$, $5_{4}$, $8_{6}$, $20_{1}$ \\
32 & $0_{89}$, $1_{12}$, $2_{16}$, $4_{4}$, $5_{4}$, $8_{2}$, $32_{1}$ \\
33 & $0_{72}$, $1_{40}$, $2_{3}$, $3_{4}$, $4_{4}$, $9_{4}$, $18_{1}$ \\
34 & $0_{85}$, $1_{14}$, $2_{17}$, $4_{7}$, $5_{1}$, $6_{2}$, $13_{1}$, $22_{1}$ \\
35 & $0_{79}$, $1_{25}$, $2_{12}$, $4_{2}$, $5_{6}$, $8_{3}$, $17_{1}$ \\
36 & $0_{86}$, $1_{14}$, $2_{17}$, $4_{4}$, $5_{2}$, $6_{2}$, $8_{2}$, $26_{1}$ \\
37 & $0_{80}$, $1_{21}$, $2_{12}$, $3_{8}$, $4_{1}$, $5_{2}$, $6_{2}$, $12_{1}$, $21_{1}$ \\
38 & $0_{74}$, $1_{32}$, $2_{8}$, $3_{3}$, $4_{5}$, $7_{5}$, $16_{1}$ \\
39 & $0_{77}$, $1_{22}$, $2_{16}$, $3_{5}$, $4_{2}$, $7_{5}$, $16_{1}$ \\
40 & $0_{70}$, $1_{36}$, $2_{7}$, $3_{7}$, $6_{7}$, $15_{1}$ \\
41 & $0_{78}$, $1_{22}$, $2_{14}$, $3_{5}$, $4_{3}$, $5_{4}$, $11_{1}$, $20_{1}$ \\
42 & $0_{74}$, $1_{26}$, $2_{15}$, $3_{5}$, $6_{7}$, $15_{1}$ \\
43 & $0_{84}$, $1_{8}$, $2_{21}$, $3_{7}$, $6_{7}$, $15_{1}$ \\
44 & $0_{78}$, $1_{24}$, $2_{3}$, $3_{15}$, $4_{6}$, $10_{1}$, $19_{1}$ \\
45 & $0_{83}$, $1_{22}$, $3_{10}$, $4_{10}$, $6_{2}$, $24_{1}$ \\
\end{tabular}}
\end{ruledtabular}
\end{table}


\begin{table*}[t]
\caption{\label{TableV}Invariants for the eight-qubit graph states.
Notation: value$_{\rm multiplicity}$. The numeration of the classes
is the one in \cite{CLMP08}.}
\begin{ruledtabular}
{\begin{tabular}{cccc} No. & Invariants & No. & Invariants \\
\hline
46 & $0_{128}$, $1_{127}$, $129_{1}$ & 97 & $0_{163}$, $1_{44}$, $2_{17}$, $3_{14}$, $4_{5}$, $5_{4}$, $7_{2}$, $10_{6}$, $22_{1}$ \\
47 & $0_{158}$, $1_{63}$, $2_{32}$, $32_{2}$, $65_{1}$ & 98 & $0_{157}$, $1_{55}$, $2_{17}$, $3_{9}$, $4_{4}$, $5_{3}$, $6_{3}$, $7_{1}$, $9_{4}$, $12_{2}$, $24_{1}$ \\
48 & $0_{173}$, $1_{64}$, $4_{15}$, $16_{3}$, $84_{1}$ & 99 & $0_{165}$, $1_{40}$, $2_{18}$, $3_{17}$, $4_{4}$, $5_{2}$, $6_{2}$, $7_{1}$, $9_{4}$, $12_{2}$, $24_{1}$ \\
49 & $0_{158}$, $1_{62}$, $2_{31}$, $16_{1}$, $17_{2}$, $32_{1}$, $50_{1}$ & 100 & $0_{152}$, $1_{59}$, $2_{16}$, $3_{8}$, $4_{12}$, $9_{8}$, $21_{1}$ \\
50 & $0_{176}$, $1_{63}$, $8_{16}$, $65_{1}$ & 101 & $0_{168}$, $1_{58}$, $4_{18}$, $8_{3}$, $9_{6}$, $12_{2}$, $24_{1}$ \\
51 & $0_{176}$, $1_{56}$, $4_{7}$, $5_{8}$, $8_{7}$, $32_{1}$, $44_{1}$ & 102 & $0_{177}$, $1_{26}$, $2_{26}$, $3_{4}$, $4_{11}$, $5_{2}$, $6_{2}$, $8_{6}$, $20_{1}$, $32_{1}$ \\
52 & $0_{192}$, $1_{24}$, $2_{16}$, $4_{8}$, $5_{7}$, $8_{4}$, $16_{4}$, $37_{1}$ & 103 & $0_{174}$, $1_{20}$, $2_{40}$, $3_{9}$, $6_{4}$, $7_{2}$, $8_{2}$, $12_{4}$, $27_{1}$ \\
53 & $0_{180}$, $1_{30}$, $2_{30}$, $4_{8}$, $8_{2}$, $10_{2}$, $16_{2}$, $17_{1}$, $49_{1}$ & 104 & $0_{200}$, $1_{21}$, $4_{24}$, $5_{6}$, $13_{4}$, $57_{1}$ \\
54 & $0_{163}$, $1_{54}$, $2_{22}$, $3_{8}$, $8_{4}$, $9_{2}$, $18_{1}$, $24_{1}$, $42_{1}$ & 105 & $0_{159}$, $1_{58}$, $2_{15}$, $3_{4}$, $4_{12}$, $9_{2}$, $10_{4}$, $16_{1}$, $34_{1}$ \\
55 & $0_{185}$, $1_{32}$, $2_{16}$, $4_{13}$, $8_{7}$, $20_{2}$, $44_{1}$ & 106 & $0_{193}$, $1_{19}$, $2_{15}$, $3_{12}$, $6_{12}$, $9_{1}$, $12_{3}$, $54_{1}$ \\
56 & $0_{181}$, $1_{30}$, $2_{23}$, $4_{6}$, $6_{7}$, $8_{3}$, $9_{2}$, $12_{2}$, $18_{1}$, $30_{1}$ & 107 & $0_{196}$, $1_{9}$, $2_{24}$, $4_{12}$, $5_{4}$, $8_{8}$, $13_{2}$, $41_{1}$ \\
57 & $0_{191}$, $1_{32}$, $2_{9}$, $4_{16}$, $10_{6}$, $16_{1}$, $66_{1}$ & 108 & $0_{180}$, $1_{26}$, $2_{26}$, $3_{4}$, $4_{4}$, $6_{10}$, $9_{2}$, $12_{3}$, $36_{1}$ \\
58 & $0_{176}$, $1_{49}$, $4_{14}$, $5_{14}$, $20_{2}$, $41_{1}$ & 109 & $0_{164}$, $1_{40}$, $2_{28}$, $3_{2}$, $4_{8}$, $5_{4}$, $6_{1}$, $7_{2}$, $10_{6}$, $22_{1}$ \\
59 & $0_{183}$, $1_{28}$, $2_{25}$, $4_{6}$, $5_{2}$, $6_{3}$, $8_{2}$, $10_{2}$, $13_{2}$, $14_{1}$, $16_{1}$, $34_{1}$ & 110 & $0_{174}$, $1_{32}$, $2_{22}$, $3_{13}$, $4_{4}$, $7_{1}$, $8_{2}$, $10_{6}$, $11_{1}$, $31_{1}$ \\
60 & $0_{179}$, $1_{32}$, $2_{25}$, $4_{9}$, $6_{3}$, $8_{3}$, $10_{1}$, $14_{2}$, $20_{1}$, $38_{1}$ & 111 & $0_{166}$, $1_{40}$, $2_{22}$, $3_{9}$, $4_{9}$, $5_{1}$, $6_{1}$, $7_{4}$, $10_{1}$, $13_{1}$, $16_{1}$, $31_{1}$ \\
61 & $0_{186}$, $1_{24}$, $2_{20}$, $4_{10}$, $5_{8}$, $8_{2}$, $10_{4}$, $16_{1}$, $40_{1}$ & 112 & $0_{168}$, $1_{31}$, $2_{32}$, $3_{9}$, $4_{8}$, $5_{1}$, $7_{2}$, $13_{4}$, $31_{1}$ \\
62 & $0_{169}$, $1_{46}$, $2_{15}$, $3_{7}$, $4_{5}$, $5_{7}$, $8_{1}$, $9_{2}$, $12_{1}$, $15_{1}$, $18_{1}$, $33_{1}$ & 113 & $0_{161}$, $1_{46}$, $2_{21}$, $3_{10}$, $4_{4}$, $5_{6}$, $6_{1}$, $8_{3}$, $11_{2}$, $14_{1}$, $26_{1}$ \\
63 & $0_{175}$, $1_{27}$, $2_{31}$, $3_{4}$, $4_{6}$, $6_{2}$, $8_{7}$, $9_{1}$, $14_{2}$, $26_{1}$ & 114 & $0_{158}$, $1_{51}$, $2_{20}$, $3_{12}$, $4_{2}$, $5_{3}$, $6_{1}$, $7_{2}$, $8_{2}$, $11_{4}$, $26_{1}$ \\
64 & $0_{200}$, $1_{8}$, $2_{14}$, $4_{18}$, $5_{6}$, $8_{6}$, $13_{1}$, $14_{2}$, $29_{1}$ & 115 & $0_{164}$, $1_{40}$, $2_{28}$, $3_{2}$, $4_{7}$, $5_{6}$, $8_{7}$, $14_{1}$, $26_{1}$ \\
65 & $0_{188}$, $1_{13}$, $2_{28}$, $4_{16}$, $6_{4}$, $9_{2}$, $12_{4}$, $33_{1}$ & 116 & $0_{161}$, $1_{38}$, $2_{37}$, $3_{7}$, $4_{2}$, $6_{1}$, $7_{3}$, $10_{6}$, $28_{1}$ \\
66 & $0_{181}$, $1_{20}$, $2_{26}$, $3_{8}$, $4_{8}$, $6_{3}$, $7_{4}$, $8_{1}$, $10_{3}$, $16_{1}$, $28_{1}$ & 117 & $0_{161}$, $1_{43}$, $2_{23}$, $3_{14}$, $4_{4}$, $5_{3}$, $6_{1}$, $7_{2}$, $10_{2}$, $13_{2}$, $28_{1}$ \\
67 & $0_{179}$, $1_{24}$, $2_{26}$, $3_{4}$, $4_{8}$, $5_{2}$, $6_{8}$, $9_{2}$, $12_{1}$, $18_{1}$, $30_{1}$ & 118 & $0_{155}$, $1_{55}$, $2_{12}$, $3_{16}$, $4_{9}$, $9_{8}$, $21_{1}$ \\
68 & $0_{170}$, $1_{35}$, $2_{20}$, $3_{12}$, $4_{7}$, $5_{2}$, $7_{4}$, $8_{1}$, $10_{2}$, $13_{2}$, $25_{1}$ & 119 & $0_{152}$, $1_{59}$, $2_{16}$, $3_{10}$, $4_{9}$, $6_{1}$, $8_{6}$, $11_{2}$, $23_{1}$ \\
69 & $0_{180}$, $1_{54}$, $4_{8}$, $5_{7}$, $16_{4}$, $17_{2}$, $37_{1}$ & 120 & $0_{160}$, $1_{42}$, $2_{29}$, $3_{3}$, $4_{12}$, $6_{1}$, $8_{6}$, $11_{2}$, $23_{1}$ \\
70 & $0_{176}$, $1_{62}$, $8_{14}$, $16_{1}$, $17_{2}$, $32_{1}$ & 121 & $0_{192}$, $1_{25}$, $4_{24}$, $5_{6}$, $8_{8}$, $41_{1}$ \\
71 & $0_{188}$, $1_{22}$, $2_{32}$, $5_{7}$, $8_{4}$, $17_{2}$, $69_{1}$ & 122 & $0_{176}$, $1_{24}$, $2_{24}$, $3_{6}$, $4_{16}$, $6_{1}$, $7_{2}$, $10_{6}$, $22_{1}$ \\
72 & $0_{148}$, $1_{84}$, $2_{8}$, $3_{7}$, $8_{4}$, $17_{4}$, $35_{1}$ & 123 & $0_{190}$, $1_{28}$, $2_{12}$, $3_{1}$, $5_{16}$, $8_{6}$, $11_{2}$, $51_{1}$ \\
73 & $0_{185}$, $1_{32}$, $2_{15}$, $4_{12}$, $8_{10}$, $16_{1}$, $50_{1}$ & 124 & $0_{200}$, $1_{5}$, $2_{32}$, $5_{6}$, $8_{8}$, $13_{4}$, $41_{1}$ \\
74 & $0_{178}$, $1_{30}$, $2_{26}$, $4_{9}$, $6_{6}$, $8_{3}$, $9_{2}$, $24_{1}$, $36_{1}$ & 125 & $0_{169}$, $1_{35}$, $2_{28}$, $3_{4}$, $4_{4}$, $5_{6}$, $6_{4}$, $8_{1}$, $9_{2}$, $12_{2}$, $33_{1}$ \\
75 & $0_{166}$, $1_{54}$, $2_{14}$, $4_{6}$, $5_{7}$, $8_{4}$, $9_{1}$, $14_{2}$, $17_{1}$, $29_{1}$ & 126 & $0_{170}$, $1_{44}$, $2_{14}$, $3_{6}$, $5_{12}$, $8_{6}$, $10_{1}$, $11_{2}$, $26_{1}$ \\
76 & $0_{188}$, $1_{26}$, $2_{14}$, $4_{14}$, $5_{2}$, $8_{6}$, $9_{2}$, $13_{1}$, $14_{2}$, $29_{1}$ & 127 & $0_{161}$, $1_{48}$, $2_{19}$, $3_{6}$, $4_{9}$, $5_{4}$, $7_{6}$, $10_{1}$, $16_{1}$, $28_{1}$ \\
77 & $0_{186}$, $1_{28}$, $2_{18}$, $4_{12}$, $5_{4}$, $10_{4}$, $14_{2}$, $16_{1}$, $40_{1}$ & 128 & $0_{161}$, $1_{42}$, $2_{33}$, $3_{3}$, $4_{6}$, $6_{1}$, $7_{3}$, $10_{6}$, $28_{1}$ \\
78 & $0_{191}$, $1_{24}$, $2_{22}$, $4_{1}$, $5_{8}$, $8_{7}$, $14_{2}$, $60_{1}$ & 129 & $0_{160}$, $1_{50}$, $2_{18}$, $3_{8}$, $4_{9}$, $6_{3}$, $7_{2}$, $9_{4}$, $12_{1}$, $30_{1}$ \\
79 & $0_{178}$, $1_{32}$, $2_{25}$, $4_{7}$, $6_{6}$, $8_{4}$, $12_{3}$, $42_{1}$ & 130 & $0_{156}$, $1_{52}$, $2_{19}$, $3_{9}$, $4_{10}$, $6_{1}$, $8_{6}$, $11_{2}$, $23_{1}$ \\
80 & $0_{166}$, $1_{49}$, $2_{15}$, $3_{6}$, $4_{8}$, $5_{6}$, $8_{1}$, $9_{1}$, $11_{1}$, $14_{1}$, $20_{1}$, $35_{1}$ & 131 & $0_{152}$, $1_{59}$, $2_{16}$, $3_{12}$, $4_{6}$, $6_{2}$, $7_{4}$, $10_{4}$, $25_{1}$ \\
81 & $0_{156}$, $1_{70}$, $2_{3}$, $3_{4}$, $4_{15}$, $9_{2}$, $10_{1}$, $13_{4}$, $28_{1}$ & 132 & $0_{156}$, $1_{52}$, $2_{16}$, $3_{13}$, $4_{10}$, $7_{5}$, $10_{2}$, $13_{1}$, $25_{1}$ \\
82 & $0_{179}$, $1_{27}$, $2_{27}$, $3_{4}$, $4_{6}$, $6_{4}$, $8_{2}$, $9_{1}$, $12_{5}$, $30_{1}$ & 133 & $0_{148}$, $1_{69}$, $2_{12}$, $3_{2}$, $4_{16}$, $9_{8}$, $21_{1}$ \\
83 & $0_{179}$, $1_{24}$, $2_{26}$, $3_{4}$, $4_{10}$, $5_{2}$, $6_{4}$, $8_{2}$, $9_{2}$, $12_{1}$, $18_{1}$, $30_{1}$ & 134 & $0_{188}$, $1_{34}$, $3_{20}$, $6_{3}$, $9_{10}$, $54_{1}$ \\
84 & $0_{165}$, $1_{49}$, $2_{14}$, $3_{6}$, $4_{10}$, $7_{6}$, $8_{1}$, $10_{2}$, $13_{2}$, $25_{1}$ & 135 & $0_{166}$, $1_{44}$, $2_{20}$, $3_{3}$, $4_{12}$, $8_{10}$, $35_{1}$ \\
85 & $0_{160}$, $1_{56}$, $2_{16}$, $3_{4}$, $4_{10}$, $5_{4}$, $8_{1}$, $14_{4}$, $32_{1}$ & 136 & $0_{191}$, $1_{30}$, $2_{3}$, $3_{3}$, $4_{12}$, $7_{15}$, $10_{1}$, $48_{1}$ \\
86 & $0_{190}$, $1_{10}$, $2_{30}$, $4_{16}$, $5_{3}$, $9_{2}$, $10_{2}$, $16_{2}$, $37_{1}$ & 137 & $0_{154}$, $1_{51}$, $2_{26}$, $3_{8}$, $4_{6}$, $6_{2}$, $7_{4}$, $10_{4}$, $25_{1}$ \\
87 & $0_{200}$, $1_{9}$, $2_{16}$, $4_{24}$, $5_{2}$, $13_{4}$, $57_{1}$ & 138 & $0_{154}$, $1_{51}$, $2_{24}$, $3_{14}$, $4_{1}$, $6_{5}$, $9_{6}$, $27_{1}$ \\
88 & $0_{176}$, $1_{28}$, $2_{16}$, $3_{14}$, $4_{12}$, $7_{1}$, $8_{4}$, $11_{4}$, $23_{1}$ & 139 & $0_{183}$, $1_{12}$, $2_{31}$, $3_{10}$, $5_{10}$, $6_{6}$, $12_{3}$, $30_{1}$ \\
89 & $0_{174}$, $1_{30}$, $2_{32}$, $4_{4}$, $5_{2}$, $6_{6}$, $8_{5}$, $14_{2}$, $32_{1}$ & 140 & $0_{160}$, $1_{36}$, $2_{34}$, $3_{9}$, $4_{8}$, $6_{4}$, $9_{2}$, $12_{2}$, $27_{1}$ \\
90 & $0_{175}$, $1_{24}$, $2_{33}$, $3_{4}$, $4_{10}$, $6_{2}$, $7_{4}$, $8_{1}$, $16_{2}$, $34_{1}$ & 141 & $0_{212}$, $1_{1}$, $3_{14}$, $6_{28}$, $45_{1}$ \\
91 & $0_{168}$, $1_{28}$, $2_{44}$, $3_{1}$, $4_{2}$, $6_{4}$, $7_{2}$, $8_{2}$, $12_{4}$, $27_{1}$ & 142 & $0_{184}$, $1_{43}$, $6_{28}$, $45_{1}$ \\
92 & $0_{175}$, $1_{27}$, $2_{31}$, $3_{4}$, $4_{8}$, $6_{2}$, $8_{1}$, $9_{1}$, $10_{6}$, $34_{1}$ & 143 & $0_{179}$, $1_{14}$, $2_{35}$, $3_{15}$, $7_{3}$, $8_{8}$, $10_{1}$, $32_{1}$ \\
93 & $0_{170}$, $1_{33}$, $2_{26}$, $3_{9}$, $4_{4}$, $5_{2}$, $6_{6}$, $7_{1}$, $9_{1}$, $12_{2}$, $15_{1}$, $27_{1}$ & 144 & $0_{172}$, $1_{9}$, $2_{56}$, $3_{6}$, $6_{4}$, $8_{8}$, $29_{1}$ \\
94 & $0_{182}$, $1_{20}$, $2_{26}$, $3_{8}$, $4_{10}$, $5_{2}$, $8_{3}$, $10_{1}$, $13_{2}$, $16_{1}$, $34_{1}$ & 145 & $0_{188}$, $1_{37}$, $3_{2}$, $6_{28}$, $45_{1}$ \\
95 & $0_{164}$, $1_{41}$, $2_{28}$, $3_{6}$, $4_{4}$, $5_{4}$, $6_{2}$, $7_{2}$, $11_{2}$, $14_{2}$, $29_{1}$ & 146 & $0_{164}$, $1_{21}$, $2_{56}$, $3_{2}$, $6_{4}$, $8_{8}$, $29_{1}$ \\
96 & $0_{167}$, $1_{40}$, $2_{23}$, $3_{7}$, $4_{5}$, $5_{5}$, $7_{1}$, $8_{4}$, $11_{2}$, $14_{1}$, $29_{1}$ & & \\
\end{tabular}}
\end{ruledtabular}
\end{table*}


\end{document}